\documentclass{elsart}

\usepackage{amsmath,amsfonts,amssymb,mathrsfs,latexsym}
\usepackage{graphics,graphicx}

\def\dfr{\mathrm d}

\newcommand{\xe}{$^{129}$Xe}

\begin{document}
\begin{frontmatter}
\title{Toward A Quantitative Understanding of Gas Exchange in the Lung}
\author{Yulin V. Chang}
\ead{yc3z@virginia.edu}
\address{Department of Mechanical Engineering and Material Science,\\
Washington University,\\
St. Louis, MO 63130, USA}
\end{frontmatter}

\newpage

\section*{abstract}
In this work we present a mathematical framework that quantifies the gas-exchange
processes in the lung. The theory is based on the solution of the one-dimensional
diffusion equation on a simplified model of
lung septum. Gases dissolved into different compartments of the lung
are all treated separately with physiologically important parameters. The model can
be applied in magnetic resonance of hyperpolarized xenon for quantification of
lung parameters such as surface-to-volume ratio and the air-blood
barrier thickness. In general this model provides a description of a broad range of
biological exchange processes that are driven by diffusion.

{\it key words}: gas exchange, air-blood barrier, surface-to-volume ratio,
hematocrit, capillary transit time, hyperpolarized xenon

\newpage
\section*{Introduction}
Gas exchange is the essential process that happens in the lung. However, quantifying
gas exchange has not been an easy task, mainly because of the complicated structure of
the lung. In general, the human lung can be seen as a porous media with capillary blood
flows inside, which mainly consists of the red blood cells (RBCs) and plasma. The blood
flows are separated from the air space by several layers of tissue, including
epithelium, endothelium and interstitium \cite{Weibel:1964}. These
tissues together are called the {\em air-blood barrier}. At equilibrium gas molecules
in the alveolar space are under constant exchange with those dissolved into
the air-blood barrier and blood in the lung.

Gas-exchange is the result of diffusion. For most gas species their concentrations are
much higher in the alveoli than in the parenchyma they dissolve into (depending on the
corresponding Ostwald solubility), and equilibrium is established once the tissue
and blood are saturated with the gases. In many situations, due to the different chemical
environment the dissolved gas molecules experience, they can usually be distinguished
from the free gas, which allows selective manipulation (e.g., RF saturation) of either
state of the gas. Such manipulations usually lead to a broken equilibrium between the
dissolved and free gases, and because gas exchange will eventually re-establish the
equilibrium, one can use this technique to quantify the gas-exchange processes
\cite{Ruppert:2000a}.

The current work is largely motivated by the recent development of magnetic resonance
(MR) of hyperpolarized \xe  (HXe) \cite{Albert:1994}. HXe makes an ideal contrast agent for
quantifying gas exchange in the lung because of its large chemical shift when dissolved
into lung tissue and blood plasma, both at 197 ppm\cite{Chang:2009a} relative to the
frequency of free xenon. In blood, xenon also dissolves into the RBCs and binds hemoglobin,
which gives rise to yet another unique peak at 217 ppm \cite{Chang:2009a}. A simplified
model of lung septum with the source of each dissolved-xenon peak identified is shown
in Fig. \ref{fig:septum}. Because of the large chemical shifts of the dissolved xenon in
the lung, one can use the technique of chemical shift saturation recovery (CSSR)
\cite{Patz:2008a} to selectively saturate the dissolved xenon and then monitor the
recovery of the xenon signals at both 197 and 217 ppm as functions of exchange time,
which we call the xenon uptake dynamics.

The growth curves of the dissolved-xenon signals carry important information of the lung.
Several groups have developed theories of the uptake dynamics using important
lung parameters \cite{Mansson:2003,Patz:2008c}. However, due to either limited range
of exchange times or limited field strengths that were unable to distinguish the two
peaks of dissolved xenon, the previously developed theories were not fully tested against
the uptake dynamics of both dissolved-xenon compartments.
In this work we presented a general theory of gas exchange in the lung with a greatly
simplified geometry of lung septum. This theory first calculates the dissolved xenon
in the air-blood barrier and in the blood separately; then, in order for it to be
used for xenon uptake dynamics, the xenon in blood is further separated into
the RBC xenon and plasma xenon. Finally the barrier (tissue) xenon and the plasma xenon
are combined for their common chemical shifts. We note that although this theory is
aimed for interpreting gas exchange in the lung, the method can also be applied to
general exchange processes in a biological system driven by diffusion.

\section*{Theory}
\begin{figure}
\begin{center}
  \includegraphics[height = 3 in]{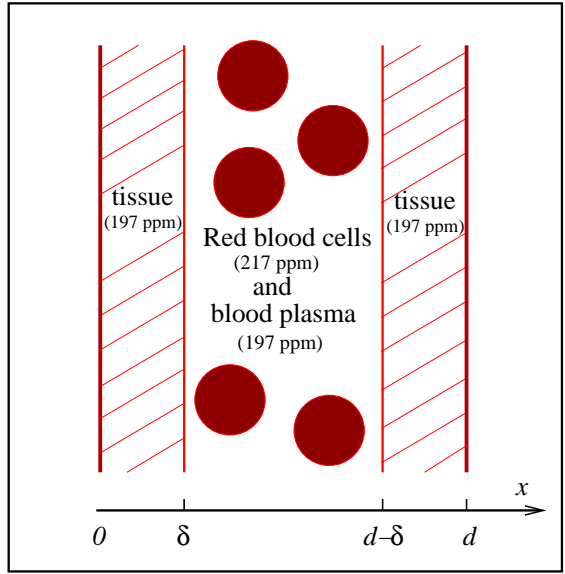}
\end{center}
  \caption{A simplified model of lung septum.}
\label{fig:septum}
\end{figure}

We begin with the 1-D diffusion equation within a simplified model of lung septum of
thickness $d$, as shown in Fig. \ref{fig:septum}. If we consider all the gas dissolved
into lung tissue as a single component, i.e., we don't distinguish between gas in tissue
and blood, then the dissolved gas density, $M_\text{d}$, as a function of the distance
$x$ from the edge of the tissue and time $t$, satisfies the diffusion equation
\begin{equation}\label{eq:1DDiffEqu}
\frac{\partial M_\text{d}}{\partial t} = D\frac{\partial^2 M_\text{d}}{\partial x^2}\,,
\end{equation}
subject to the initial condition
\begin{equation}\label{eq:1DDiffInit}
M_\text{d}(x, 0) = 0\;\;, x\in (0,d)\,,
\end{equation}
and the boundary conditions
\begin{equation}\label{eq:1DDiffBound}
M_\text{d}(0,t) = M_\text{d}(d, t) = \lambda M_\text{f}\,.
\end{equation}
In the above equations, $M_\text{f}$ is the density of free gas (at 0 ppm), $D$ is the
diffusion coefficient of dissolved gas, $d$ is the septal thickness, and $\lambda$ is
the Ostwald solubility of xenon in lung septum. We also made the
assumption that all the dissolved gas molecules that hit the wall at a certain time $t$
leave the tissue (i.e., exchange with the free gas molecules).

The complete solution of this problem is well known and can be represented by a sum
of infinite series:
\begin{equation}\label{eq:1DDiffFullSolu}
M_\text{d}(x,t) = \lambda M_\text{f}\biggl\{1-\frac{4}{\pi}\sum_{n=1}^{\infty}\frac{1}
{2n-1}\sin\frac{(2n-1)\pi x}{d}\,e^{-D\frac{(2n-1)^2\pi^2}{d^2}t}\biggr\}\,.
\end{equation}
In a real experiment the density distribution $M_\text{d}(x,t)$ can be converted
in the signal distribution $S_\text{d}(x,t)$ of the dissolved gas. $S_\text{d}(x,t)$
is proportional to the total surface area $S_\text{A}$ in the lung, and can be normalized
by the total free-gas density, $M_\text{f}V_\text{g}$, where $V_\text{g}$ is the
gas volume in the lung. If we define the {\em gas exchange time constant} $T$ as
\begin{equation}\label{eq:timeConst}
T = \frac{d^2}{\pi^2 D}\,,
\end{equation}
then Eq. \eqref{eq:1DDiffFullSolu} can be simplified to
\begin{equation}\label{eq:dissXeMagnetization}
S_\text{d}(x,t) = \lambda\frac{S_\text{A}}{V_\text{g}}\biggl\{1-\frac{4}{\pi}
\sum_{n=\text{odd}}\frac{1}{n}\sin\frac{n\pi x}{d}\,e^{-n^2t/T}\biggr\}\,.
\end{equation}

We now consider the dissolved-gas signal from each compartment in the lung. As shown in
Fig. \ref{fig:septum}, we use a simple model that assumes a layer of tissue (air-blood
barrier) of thickness $\delta$ ($\delta < d/2$) at each side of the septum. The total
signal from the tissue, denoted by $S_\text{d1}(t)$, can be calculated as the spatial
integral of $S_\text{d}$ in Eq. \eqref{eq:dissXeMagnetization} over the two regions
from $0$ to $\delta$ and from $d-\delta$ to $d$. If we let
$\mu = \lambda dS_\text{A}/V_\text{g}$ and $\kappa = \delta/d$, then
the the gas signal from the tissue compartment $S_\text{d1}$ is
\begin{equation}\label{eq:TXeSignal}
S_\text{d1}(t) = \mu\biggl\{2\kappa-\frac{8}{\pi^2}\sum_{n=\text{odd}}\frac{1}{n^2}
\bigl[1-\cos(n\pi\kappa)\bigl]e^{-n^2t/T}\biggr\}\,.
\end{equation}
The gas signal from the blood is more difficult to calculate due to the flow effect.
Assuming no flow, the gas signal from the static blood, $S_\text{d2s}(t)$, is simply
the spatial integral of $S_\text{d}$ over the region from $\delta$ to $d-\delta$:
\begin{equation}\label{eq:CapXeSignalNoFlow}
S_\text{d2s}(t) = \mu\biggl\{(1-2\kappa)-\frac{8}{\pi^2}\sum_{n=\text{odd}}
\dfrac{1}{n^2}\cos(n\pi\kappa)\:e^{-n^2t/T}\biggr\}\,.
\end{equation}

\begin{figure}
\begin{center}
\includegraphics[height=3 in]{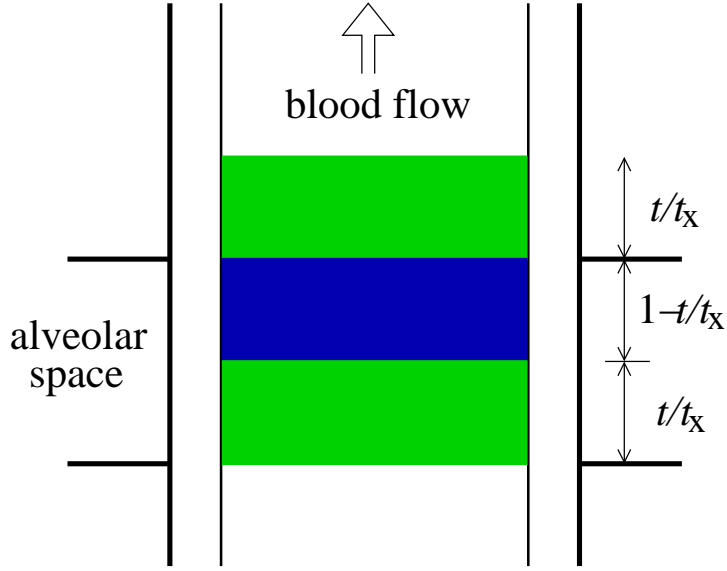}
\end{center}
\caption{Lung septum with blood flow.}
\label{fig:septumBloodFlow}
\end{figure}
To deal with the partial volume effect caused by the blood flow, we follow the method
given by Patz in \cite{Patz:2008c}. In brief, we divide the
region with dissolved gas into three sections, as shown in Fig. \ref{fig:septumBloodFlow}.
First we introduce a new quantity, the {\em pulmonary capillary transit time}, defined
as the average time an RBC spends in the gas-exchange zone (i.e., in contact with the
alveolar space shown in Fig. \ref{fig:septumBloodFlow}) in the lung, and denoted by
$t_\text{X}$.
If we let $t$ be the exchange time, and $\tau$ be the time a certain infinitesimally thin
layer of blood spends in the gas-exchange zone, then the signal contribution from the
blood gas in this thin layer is, in terms of $S_\text{d2s}$,
$S_\text{d2s}(\tau)\dfr\tau/t_X$, and therefore the total contribution of this ``partial
volume'' (the two green areas in Fig. \ref{fig:septumBloodFlow}) can be calculated with
a time integral over $S_\text{d2s}$; on the other hand,
for $t<t_\text {X}$, the region of $(1-t/t_\text{X})$ thick (blue area in Fig.
\ref{fig:septumBloodFlow}) is always in contact with the gas-exchange zone during
$t$, thus the signal with this section can still be treated as static.
Therefore the total signal is
\begin{equation}\label{eq:totalBloodSig}
\begin{split}
S_\text{d2} & = 2\int_0^t\frac{\dfr\tau}{t_\text{X}}S_\text{d2s}(\tau)
+\left(1-\frac{t}{t_\text{X}}\right)S_\text{d2s}(t)\\
& = 2\mu\biggl\{(1-2\kappa)\frac{t}{t_\text{X}}-\frac{8}{\pi^2}\frac{T}{t_\text{X}}
\sum_{n=\text{odd}}\frac{1}{n^4}\cos(n\pi\kappa)\left(1-e^{-n^2t/T}\right)\biggr\}\\
&\quad+\mu\left(1-\frac{t}{t_\text{X}}\right)
\biggl\{(1-2\kappa)-\frac{8}{\pi^2}\sum_{n=\text{odd}}
\frac{1}{n^2}\cos(n\pi\kappa)e^{-n^2t/T}\biggr\}\,,
\end{split}
\end{equation}
where I have used the integral
\begin{equation}
\int_0^te^{-n^2\tau/T}\dfr\tau = \frac{T}{n^2}(1-e^{-n^2t/T})\,.
\end{equation}

In order for this theory to be applied to dissolved-HXe uptake dynamics we need
to find expressions for the signal amplitudes at 197 ppm and 217 ppm. As shown
in Fig. \ref{fig:septum}, for humans, the HXe in tissue and blood plasma share the
same resonant frequency at 197 ppm, and HXe in the RBCs resonates at 217 ppm.
Thus if we let $\eta$ be the fraction of dissolved gas in the red blood cells,
the peak amplitude at 197 ppm, $S_\text{TP}$, combining HXe in tissue and plasma, is
\begin{equation}\label{eq:TP}
\begin{split}
S_\text{TP}(t) & = S_\text{d1}(t) + (1-\eta)S_\text{d2}(t)\\
& = \mu\biggl\{2\kappa-\frac{8}{\pi^2}\sum_{n=\text{odd}}\frac{1}{n^2}
\bigl[1-\cos(n\pi\kappa)\bigl]e^{-n^2t/T}\biggr\}\\
&\quad + 2\mu(1-\eta)\biggl\{(1-2\kappa)\frac{t}{t_\text{X}}-\frac{8}
{\pi^2}\frac{T}{t_\text{X}}\sum_{n=\text{odd}}\frac{1}{n^4}\cos(n\pi\kappa)
\left(1-e^{-n^2t/T}\right)\biggr\}\\
&\quad +\mu(1-\eta)\left(1-\frac{t}{t_\text{X}}\right)\biggl\{(1-2\kappa)
-\frac{8}{\pi^2}\sum_{n=\text{odd}}\frac{1}{n^2}\cos(n\pi\kappa)\,
e^{-n^2t/T}\biggr\}\,,
\end{split}
\end{equation}
and the peak amplitude of HXe in the RBCs at 217 ppm, $S_\text{RBC}$, is
\begin{equation}\label{eq:RBC}
\begin{split}
S_\text{RBC}(t) & = \eta S_\text{d2}(t)\\
& = 2\mu\eta\biggl\{(1-2\kappa)\frac{t}{t_\text{X}}-\frac{8}
{\pi^2}\frac{T}{t_\text{X}}\sum_{n=\text{odd}}\frac{1}{n^4}\cos(n\pi\kappa)
\left(1-e^{-n^2t/T}\right)\biggr\}\\
&\quad +\mu\eta\left(1-\frac{t}{t_\text{X}}\right)\biggl\{(1-2\kappa)
-\frac{8}{\pi^2}\sum_{n=\text{odd}}\frac{1}{n^2}\cos(n\pi\kappa)\,
e^{-n^2t/T}\biggr\}\,.
\end{split}
\end{equation}

\section*{Discussion}
\begin{figure}
\begin{center}
\includegraphics[height=3 in]{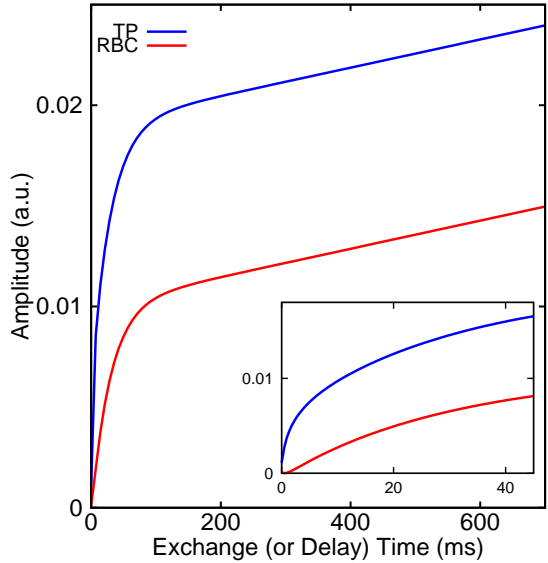}
\end{center}
\caption{Plots of $S_\text{TP}$ and $S_\text{RBC}$ in Eqs. \eqref{eq:TP} and
\eqref{eq:RBC}, respectively, for $n\le 9$, as functions of the gas-exchange time up to
700 ms. The parameters used are: $\mu = 0.03$, $\kappa = 0.15$, $\eta = 0.5$, $T = 30$
ms and $t_\text{X} = 1.5$ s. The insert show details of the curves at short exchange
times.}
\label{fig:XeCurves}
\end{figure}
Test of the theory using HXe experiments will be presented separately. Nevertheless,
insights can be gained by plotting $S_\text{TP}$ and $S_\text{RBC}$ in Eqs.
\eqref{eq:TP} and \eqref{eq:RBC}, respectively, as functions of the gas-exchange time
for $n\le 9$, as shown in Fig. \ref{fig:XeCurves}. The plots
are made using the following estimated parameters for the human lung: $\mu = 0.03$,
$\kappa = 0.15$, $\eta = 0.5$, $T = 30$ ms and $t_\text{X} = 1.5$ s. These two curves,
by comparison, are similar in shapes and characteristics with the previously published
dissolved-HXe data \cite{Mansson:2003,Driehuys:2006,Chang:2008a}, which supports the
validity of the presented theory. We would like to pointed out two interesting features of
the curves. First, for exchange time longer than 100 ms, both curves grow linearly with
time. This is because both tissue and blood are saturated with gas (HXe in this case)
at about 100 ms and the signal growth after which is purely due to blood flow; second,
although blood is never is directly contact with the free gas in alveoli, there is
still signal from RBC at the very short of exchange time. This is a result of diffusion
--- the propagation of probability density in diffusion is non-local.

The presented theory can potentially be used to fit the experimental data and extract
important parameters related to lung physiology and function. The parameters $\mu$,
$\kappa$ and $t_\text{X}$ represent, respectively, the product of septal thickness
$d$ and the surface-to-volume ratio $S_\text{A}/V_\text{g}$, the ratio between the
air-blood barrier $\delta$ and septal $d$ thicknesses, and the pulmonary capillary
transit time. $\eta$ is the partition of the RBC HXe in the blood, which can be used
to calculate the hematocrit (Hct) using
\begin{equation}\label{eq:Hct}
\text{Hct} = \frac{\eta/\lambda_\text{RBC}}{\eta/\lambda_\text{RBC}+(1-\eta)
/\lambda_\text{P}}\,,
\end{equation}
where $\lambda_\text{RBC}$ and $\lambda_\text{P}$ are the Ostwald solubilities of HXe
in the RBCs and plasma, respectively.

The current model used a fair number of simplifications, including the na\"{i}ve
geometry of the septum, undistinguished gas solubilities and diffusion coefficients
in the tissue and blood. Therefore, great improvements can be done by carefully
dealing with these factors.

\section*{Conclusions}
We have developed a theory of gas-exchange in the lung based on a simple model of
1-D gas diffusion. This theory carefully treats the dissolved gas in different
compartments of lung septum with parameters of physiological importance.
It can be used to quantify pulmonary function using the dynamics of the dissolved
hyperpolarized xenon in the lung.

\section*{Acknowledgments}
The author is financially supported by NIH grant R21EB005834 (PI: Philip V. Bayly).
I thank Dr. Kai Ruppert for helpful comments.

\bibliographystyle{unsrt}
\bibliography{/home/ychang/Documents/notes/references/ycMRI.bib}
\end{document}